\documentclass[%
 preprint, 
superscriptaddress,
 amsmath,amssymb,
 prd, aps, twocolumn,
floatfix,
]{revtex4-2}

\usepackage[top=2cm, bottom=2cm, left=2cm, right=2cm ]{geometry}
\usepackage{graphicx}% Include figure files
\usepackage{dcolumn}% Align table columns on decimal point
\usepackage{natbib}
\usepackage{hyperref}
\usepackage{textcomp}
\usepackage{natbib}
\usepackage{bm}% bold math
\usepackage{amsfonts}
\usepackage{amssymb}
\usepackage{hyperref}% add hypertext capabilities
\usepackage{siunitx}
\usepackage{caption}
\usepackage{subcaption}
\usepackage{enumitem}
\usepackage{todonotes}
\usepackage{placeins}
\usepackage{setspace}
\begin{document}

%\preprint{APS/123-QED}

\title{\textbf{First Experimental Evaluation of Stray Light Noise Modeling \\
of kilometers-long Beam Tube Arms of Gravitational Wave Detectors } 
}% 

\author{Emilia Chick}
\email{Contact author: echick@star.sr.bham.ac.uk}
\affiliation{School of Physics and Astronomy, University of Birmingham, Birmingham, United Kingdom}

\author{Timothy O'Hanlon}
%\email{Contact author: timothy.ohanlon@ligo.org }
\affiliation{LIGO Lab, California Institute of Technology, Pasadena, CA, USA}

\author{Hiro Yamamoto}
%\email{Contact author: yamamoto.h.c51c@m.isct.ac.jp}
\affiliation{Institute of Science Tokyo, Tokyo, Japan}

\author{Denis Martynov}
%\email{Contact author: dmartynov@star.sr.bham.ac.uk}
\affiliation{School of Physics and Astronomy, University of Birmingham, Birmingham, United Kingdom}

\author{Valery Frolov}
%\email{Contact author: valera.frolov@ligo.org}
\affiliation{LIGO Lab, California Institute of Technology, Pasadena, CA, USA}

\author{Anamaria Effler}
\email{Contact author: aeffler@caltech.edu}
\affiliation{LIGO Lab, California Institute of Technology, Pasadena, CA, USA}
\author{Antonios Kontos}
\email{Contact author: akontos@bard.edu}
\affiliation{Physics Program, Bard College, Annandale-on-Hudson, NY, USA}

\date{\today}% It is always \today, today,

\begin{abstract}
We present the results of the first experimental evaluation of stray light noise models for LIGO beam tube baffles. We combine the results of a testing campaign at LIGO Livingston and SIS simulations to get a model of backscatter and diffraction noise of select mid-arm baffles. Agreement to within a factor of $3-10$ is found between models and measured gravitational wave strain. Two distinct regions of well-propagating motion were identified, at approximately \SI{35}{\hertz} and \SI{90}{\hertz}, and these are visible in the strain signal as broadband features.  Peaks with $Q \approx 100-1000$ were also found in the strain signal between \SI{70}{\hertz} and \SI{100}{\hertz}. These are theorized to come from diffraction noise only, and are modeled using Q-factors measured from the strain.  These models are also in good agreement with the strain signal.  Other findings on the effect of auxiliary arm systems on stray light noise are discussed. 
\end{abstract}
\maketitle

\section{\label{sec:intro} Introduction}

Ground-based gravitational wave observatories are among the most sensitive instruments ever constructed, as they must be capable of measuring very small distortions in spacetime. They achieve this sensitivity by employing low-loss, 4-kilometer-long optical Michelson cavities held under high vacuum \cite{AdvLigo}, coupled with a thorough approach to noise identification and mitigation \cite{Soni2025, Glanzer_paper}.
%Since the first detection in 2015 \cite{GW150914}, the sensitivity of the detectors has continued to increase. The binary neutron star inspiral range \cite{finn-chernoff, chen21, senO3} increased from around $80$Mpc in observing run 1 (O1) \cite{O1O2sen} to approximately $160$Mpc to $170$Mpc in observing run 4 (O4), the most recent observing run \cite{O4sen, GWTC5}.LIGO and VIRGO measured the first binary Neutron Star merger during observing run 2 in 2017 (GW170817) \cite{GW170817}, and alongside the detection of the electromagnetic gamma-ray burst counterpart \cite{GWGRB, fermi}, this became the first multi-messenger detection of a collision event. This allowed scientists to observe the whole merger process for the first time. In order to increase the capabilities of these observatories, potentially leading to more multi-messenger events being observed, their sensitivity must continue to increase. This is planned with the LIGO A\# upgrades scheduled for 2029 \cite{postO5}, and the construction of third generation ground-based detectors like Cosmic Explorer (CE)\cite{CE-WP} and Einstein Telescope (ET) \cite{ETSC}. 

Since the first detection in 2015 \cite{GW150914}, the sensitivity of the observatories has continued to increase. The observing range of the LIGO observatories \cite{finn-chernoff, chen21, senO3} has approximately doubled between observing run 1 \cite{O1O2sen} and the most recent observing run 4 \cite{O4sen, GWTC5}. LIGO and VIRGO also measured the first binary Neutron Star merger during observing run 2 in 2017 (GW170817) \cite{GW170817}, and alongside the detection of the electromagnetic gamma-ray burst counterpart \cite{GWGRB, fermi}, this became the first multi-messenger detection of a collision event. This allowed scientists to observe the whole merger process for the first time. In order to increase the capabilities of these observatories, potentially leading to more multi-messenger events being observed, their sensitivity must continue to increase. This is planned with the LIGO A\# upgrades scheduled for 2029 \cite{postO5}, and the construction of third generation ground-based detectors like Cosmic Explorer (CE)\cite{CE-WP} and Einstein Telescope (ET) \cite{ETSC}. 

Achieving the higher sensitivity needed to increase the rate of merger detection requires a detailed understanding of the various noise sources that may inhibit the detector. As the strain sensitivity of the LIGO observatories in particular increases, the stray light noise from the beam tube itself and beam tube baffles may become a limiting noise source. Due to the technical difficulty in accessing the multi km-long beam tube interior after construction, the LIGO design took great care to model and mitigate scattered light noise from the surface of the beam tube \cite{BT_design}. LIGO's \SI{4}{\kilo\meter} beam tube \cite{AdvLigo} has circa 200 serrated, conical baffles inside each of the long arms, which shield the wall from any ray originating from a test-mass mirror \cite{BT_design}.  Future proposed detectors, such as CE and ET are following suit and addressing this design challenge before other efforts \cite{CE_Design, CE_sus_Baff, ScattLightYam, Andres_Carcasona_2025}. 

Due to the conservative baffle strategy of the current LIGO detectors \cite{BT_design} noise originating from the beam tube has been negligible up to this point. Therefore, much of the previous work on stray light control and scatter has focused on noise sources surrounding key optics. For example, wide-angle scatter from point defects like dust on the mirror surface was raised as a source of stray light noise during the advanced LIGO (aligo) upgrade \cite{WideAngle}. Work has been done in mitigating this via the installation of the cage baffles at LIGO Livingston \cite{O4scatter}, and reaction-chain tracking during observing run 3 \cite{soni2021,GWTC3}. Scatter from the many smaller elements within the beam path have also been considered as a source of stray light noise during the aligo upgrades \cite{o2o3detchar, Capote2025, Soni2025}. This is primarily concerned with defects or specular reflections from the components themselves as opposed to the structures within the vacuum envelopes. One similar analysis into beam tube baffle backscatter noise was performed by Fritschel and Yamamoto in 2013 \cite{Fritschel}, when addressing the effects of coated surface variations on aligo end test masses (ETMs). ETMs labelled number 7 and number 9 were coated at the same time using the planetary coating process, and were found to have a surface ripple \cite{CoatRip}. The ripple was scattering light into an annulus around the beam axis which would intersect the beam tube baffles, and so the projected increase in strain noise was investigated. However, this analysis did not include an extensive injection campaign as it was beyond the scope of the work.

Prior to the measurements described in this paper, no experimental data existed to verify stray light models. Recent O4 level sensitivities \cite{Capote2025} were needed to be able to measure a signal in the GW channel, despite the ability to locally increase the beam tube motion by a few orders of magnitude. Therefore, it is imperative we validate the current stray light models in order to best inform the baffle strategy for increasingly sensitive, and more complex, next generation detectors.

In this work, we have found two main mechanisms for inducing scattered light noise that is visible above the background sensitivity. Firstly, due to the mechanical boundary conditions of the beam tube, only a few frequencies can travel and increase the motion over several km of the beam tube. For these specific frequencies, we can induce motion in all the baffles along the beam tube, which creates backscatter noise. Secondly, our results suggest that some of the baffles have high resonances with Q factors in the several hundreds to a few thousand, and so they amplify the motion of a particular one or few nearby baffles by this resonance. This can produce diffraction noise high enough to be measured. These measurements allow us to make predictions of the current beam tube scattered light noise limits and provide validated models for the design of future gravitational-wave detectors.

Section \ref{sec:model} provides a brief summary of the mathematical framework used to model stray light from beam tube baffles, namely backscatter and diffraction noise. Section \ref{sec:meas} explains the measurements taken during the testing campaign at LIGO Livingston in 2024. It also details how the motion of the beam tube is calculated and used in our models. Finally, section \ref{sec:results} shows the final models compared to our measured injections, and the conclusions thereof.

\section{Noise Modeling}\label{sec:model}

Modeling stray-light noise in the beam tube involves: a) accurately propagating the stray light from the cavity mirrors to the beam tube walls and baffles, and b) understanding the physical mechanisms that re-couple the stray field into the main cavity after it interacts with these surfaces. Noise in the cavity translates to noise in the strain signal. For the field propagation, we use the FFT-based code Static Interferometer Simulation (SIS) ~\cite{SIS}, which incorporates measured surface maps of the cavity mirrors~\cite{coreoptics} and realistically propagates the field that scatters out of the cavity's TEM00 mode and onto the beam tube surfaces and baffles. The simulation's accuracy in propagating higher order fields is limited by the FFT resolution and the paraxial approximation, but it is sufficient for estimating the power on the beam tube baffles. The coupling of the scattered light back into the cavity is calculated using the formalism outlined in the work of Flanagan and Thorne ~\cite{T940063}. The two main mechanisms that we consider and are relevant to our measurements are the backscattering from the tube baffle surface, and  diffraction of the field from the tube baffle aperture. We describe the mechanisms in the following sections, and present the relevant formulas without proof. The reader is directed to~\cite{T940063, T950101, T1100646} for a detailed treatment.

There are other potential sources of stray light noise couplings from within the arm tube, such as backscatter from the tube walls \cite{Vinet1997} and diffraction-aided specular reflection \cite{diff_aided}, but they were negligible given the LIGO tube baffle coverage. It is also noted that there are many scattered light sources that originate from in-vacuum structures at the vertex and ends of the interferometer which may have a greater impact on the sensitivity of the detector. This is beyond the scope of this campaign, and is left for future work to explore.

\subsection{Backscatter Noise from beam tube baffles}

The LIGO baffles are oriented such that the smaller, serrated inner radius of the conical baffle faces away from the nearest test mass, so the direction they face flips at the middle of the beamtube. They are angled at $55^\circ$ relative to the beam tube surface normal, which is $35^\circ$ to the beam tube wall \cite{BT_design}. This is so that any stray light directed to them is either trapped between them and the beam tube wall, or redirected further down the beam tube at oblique angles. This makes the recombination of the light back into the cavity after multiple reflections unlikely. While most of the power will follow the specular path, there is some that will scatter backward in the direction of the cavity mirror. The backscattered light then hits the mirror and has a chance of scattering back into the TEM00 mode of the cavity. Any stray light that follows this path will have phase fluctuations imprinted upon it by the baffle surface motion. The beam tube baffles have very little common motion with the optical system as they have a direct connection to the ground. On the other hand, the aligo cavity mirrors have tailored infrastructure providing up to seven levels of seismic isolation.\cite{seismic}. 

The recombination of backscatter light into the cavity mode induces two types of noise: phase noise and radiation pressure noise.  This radiation pressure noise arises when we consider the cavity power fluctuations as a result of the backscatter from the baffles, and the induced motion of the test masses as a result. This component is dependent on the frequency of the induced motion, and is summed together with the phase noise to give the total backscatter. A derivation of the backscatter noise is outlined more thoroughly in  \cite{Fritschel}. The strain noise induced by these processes from the i-the baffle is estimated by:

\begin{equation}
\label{eq:backscatter}
\begin{split}
h_i^2(f) &=
\frac{1}{r_i^2}\left(\lambda^2 +\left(\frac{8 \Gamma I_{mb}}{cM\pi f^2}\right)^2\right) \\
&\quad \times
\left(\frac{\lambda}{4\pi L}\right)^2
\frac{dP}{d\Omega_{bs}}
\tilde{S}_i^2(f) \\
&\quad \times
\int_{baffle_i}
\left(\frac{dP}{d\Omega_{ms}}\right)^2
\, d\Omega_{ms}.
\end{split}
\end{equation}

where $r_i$ is the distance between the baffle and mirror, $dP/d\Omega_{bs}$ is the probability of backscattering from baffle to mirror per solid angle, $\tilde{S}_i(f)$ is the amplitude spectral density of the sine of the phase shift accumulated to the baffle motion (see section \ref{sec:fw}), $dP/d\Omega_{ms}$ is the probability that the mirror scatters light towards the baffle per solid angle, calculated by SIS using measured mirror surface phase maps, $\Gamma = 13.6$ is the cavity signal gain, $I_{mb} = \SI{300}{\kilo\watt}$ is the main beam power, $c$ is the speed of light, and $M = \SI{40}{\kilo\gram}$ is the mirror mass

The term $dP/d\Omega_{bs}$ is essentially the bidirectional reflection distribution function (BRDF) \cite{BRDF_nic} of the baffle material without the cos($\theta$) geometric correction, evaluated at the backscattering angle, which in this case is $55^\circ$. LIGO uses oxidized stainless steel baffles, which have a measured BRDF value of $ \approx 10^{-2}$~str$^{-1}$ at $55^\circ$ \cite{Baffbrdf}. This work uses a conservative value of $4\times10^{-2}$~str$^{-1}$. 

The integral is performed over the full surface of the baffle, and assumes all contributions from the infinitesimally small surface areas of the baffle are summed incoherently due to surface roughness.  To get some intuition on the formula, we note that the mirror scattering probability per solid angle enters twice in the formula (as seen from the square term in the integral). The coupling mechanism requires that the light scatters out of the cavity mode and then scatters back in, and the probabilities of the two processes are linked through reciprocity \cite{T940063}. The noise contributions for different baffles and for the two beam directions are added incoherently.

%Backscatter noise is made of two components: Phase noise as light bounces off the baffles (shown in eqn. \ref{eq:backscatter}), and radiation pressure noise.  This radiation pressure noise arises when we consider the cavity power fluctuations as a result of the backscatter from the baffles, and the motion induced on the test masses as a result. This derivation is outlined more thoroughly in  \cite{Fritschel}. In summary, we also need to include the following term, denoted A. We cannot add this into the timeseries easily, due to the frequency dependence in the noise. We therefore add this noise to the phase term once we have converted it into the frequency domain. The radiation pressure strain noise Power Spectral Density (PSD) is also given below. 

%The crossover in their strain contribution is at approximately \SI{30}{\hertz} for this model, and this will shift slightly for each injection. This is in the frequency regions of the injections, so we must consider both components. Figure \ref{fig:bsvrad} shows the two component's contribution to the strain noise for an injection of \SI{20}{\hertz} to \SI{40}{\hertz} at baffle X107. We see the two terms cross over at \SI{28.53}{\hertz}, in the injected frequency band.

\subsection{Diffraction Noise from beam tube baffles}

The cavity TEM00 mode is inevitably disrupted by the finite mirror size ($R_{TM}=0.17$m) and beam tube baffles aperture ($R_{b}=0.26$m) \cite{AdvLigo}. The transverse motion of the beam tube baffles causes a time varying diffracted field to propagate down the beam tube and imprint phase noise on the cavity mode. The result of the analytical treatment of this diffraction noise is presented in~\cite{T950101}, but in this work we follow the numerical treatment implemented in SIS and described in~\cite{SIS}. 

Specifically, SIS calculates the relative change of the field amplitude $\delta a_{00}$ that resonates in the cavity, when the i-th baffle shifts transversely (perpendicular to the beam axis) by some small distance $+\delta$ and $-\delta$. Assuming that the field change is small, its imaginary part is related to a phase shift and consequently an equivalent strain $h$ by:
\begin{equation}
    \label{eq:diffraction}
    h_i(f) = \frac{\lambda}{4\pi L}\frac{\Im(\delta a_{00})_i}{2\delta}\tilde{X}_i(f),
\end{equation}
where $\tilde{X}_i(f)$ is the amplitude spectral density of the i-th baffle motion in a transverse direction (horizontal or vertical). Noise contributions from different baffles and for the two beam directions are added incoherently. A more detailed description of this numerical estimation of diffraction noise will be presented in a future publication, and goes beyond the scope of this work. 

\subsection{Frequency Upconversion due to Fringe wrapping}\label{sec:fw}

The spectrum of the baffle motion enters equations~\ref{eq:backscatter} and~\ref{eq:diffraction} above in slightly different ways. In the case of backscatter, the relevant quantity is the spectrum of the sine of the phase shift experienced by a scattered field, that makes the round trip between the mirror and the baffle. Specifically, given a baffle motion time series $x_b(t)$, we calculate:
\begin{align}
    S(t) &= sin(4\pi x_b(t)/\lambda),
    \label{eq:fringe}
\end{align}
from which we calculate $\tilde{S}(f)$. If the motion of the baffle is small compared to the laser wavelength, $x_{rms}\ll\lambda$, then $sin(4\pi x_b(t)/\lambda)\approx 4\pi x_b(t)/\lambda$, and therefore the relation between the noise spectrum and the baffle motion spectrum is linear. However, since beam tube baffles are not seismically isolated, $x_{rms}$ can easily be greater than $\lambda$, and the relation between motion and backscatter noise spectra is non-linear \cite{Soni_2021}. This is a phenomenon which is often referred to as \emph{fringe wrapping} or \emph{phase wrapping}, and leads to frequency upconversion. For example, a high-Q resonance at some frequency in the beam tube baffle motion spectrum will appear very different in the backscatter noise spectrum; it is often flatter and appears more like a broadband feature compared to a sharp peak. This phenomenon will be crucial in correctly modeling the measured noise spectra, as will be discussed in a later section.

On the other hand, diffraction noise is linear with respect to the beam tube baffle spectra for motions that are much greater than the laser wavelength. A good estimation for the validity condition of linearity is that the motion needs to be much smaller than the spacing between Fresnel zones. For a \SI{4}{\kilo\meter} beam tube, at 0.5 meters transverse distance, and a 1~$\mu$m wavelength the Fresnel zone spacing is  $\sim 0.2-1.0$~mm, depending on the location of the baffle. Therefore, for $x_{rms}\ll0.2$~mm, we do not expect any upconversion in the noise spectrum from diffraction noise. 

%\begin{figure*}[t]
   % \centering
    %\includegraphics[width=0.6\textwidth]{Images/Plot_phaseshift_vs_shift_vs_baffle_LIGO.pdf}
    %\caption{Numerical test of linearity of the diffraction noise coupling. The simulations show that the coupling (y-axis) is independent of the baffle transverse shift (x-axis) for values $\delta<0.5$mm for a baffle at 2km, and $\delta<0.2$mm for a baffle at 100m. The vertical lines show the corresponding limits calculated from the Fresnel zone estimation.}
   % \label{fig:linearity}
   
%\end{figure*}

%\anamaria{I feel that the modeling for our measurements should go in the measurements section after we describe the data and maybe this section should be more of a review of current state of models and their uncertainties and/or some slightly more in-depth introduction about which scattered light effects matter, etc? what do others think?}
%\millie{I think I agree. I think it will flow better if our modeling goes after we have the measurements, so maybe this section should focus on the backscatter and diffraction components, and maybe how SIS calculates these things?}
%\antonios{Yes, I will include here the general formulas and what kind of noise we are considering, and what kind of noise we are not because it is negligible.}

\section{Measurements}\label{sec:meas}

Measurements of baffle scatter were made at LIGO Livingston in March - June 2024. The aim of the tests were to inject vibrations into the beam tube at various positions or key baffles, using both an electromagnetic shaker and connector rod \cite{Austin_thesis} and acoustic equipment. The aim was to see an impact of these injections in the differential arm length signal (DARM), from which we derive strain. 

\begin{figure*}[!t]
    \centering
    \includegraphics[width=0.9\textwidth]{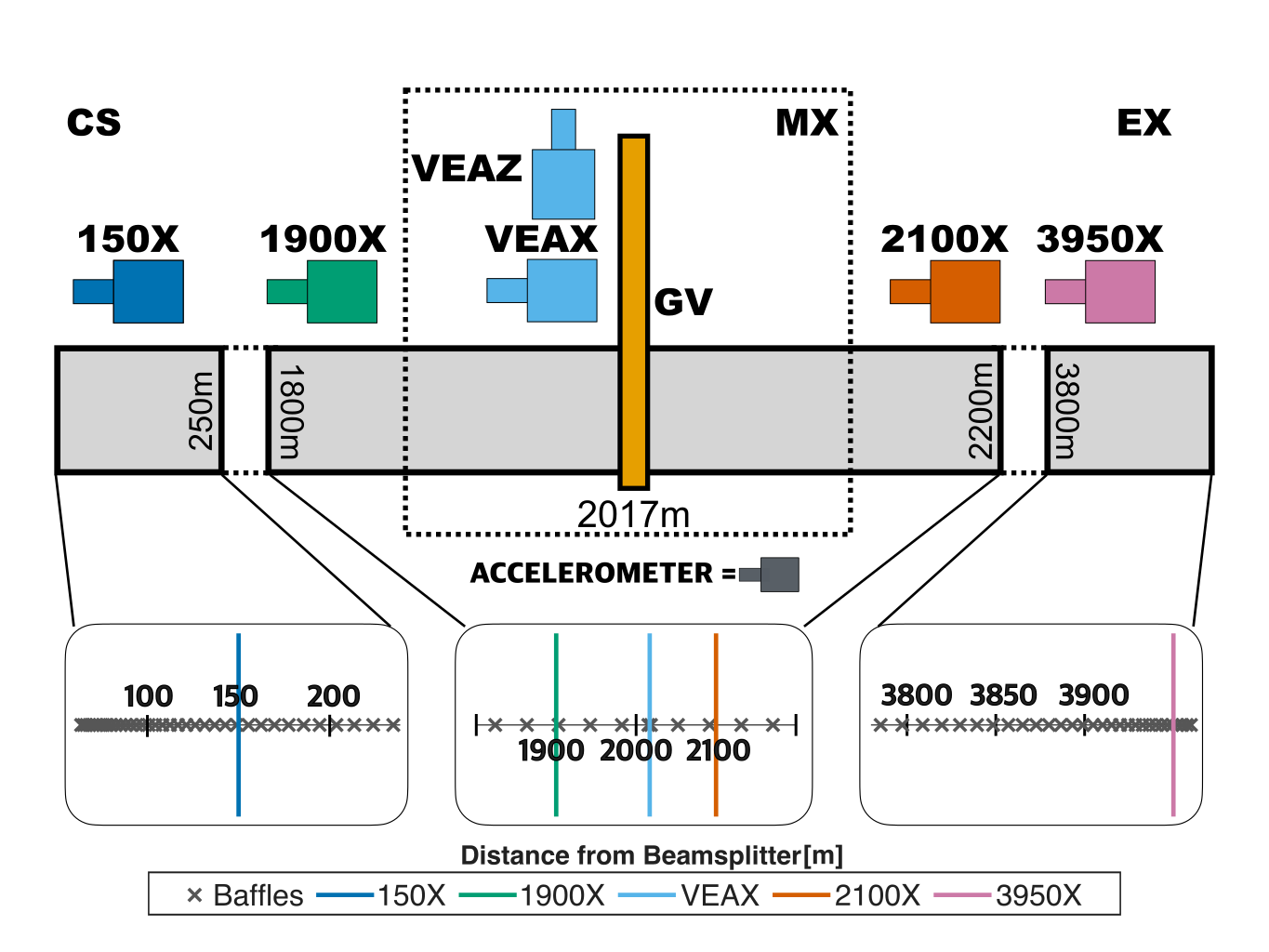}
    \caption{The schematic show a cartoon drawing (top) of the location of the sensors used for the measurements on the LIGO Livingston X arm; the placement is identical on the Y arm. The bottom plots are zoom ins of the 3 areas that are instrumented to show the beam tube baffle distribution with respect to our sensors.}
    %\anamaria{added this fig, feedback pls}
    %\millie{Looks good to me! I was going to use baffdiags.png but this one's probably better!}
    %\antonios{I think the layout and information it relays is exactly right. But they y-axis is a bit awkward and the text probably too small, the journal may complain. I will take a crack at it based on what you have.}}
    \label{fig:AccDiagram}
\end{figure*}

Injection sites were constrained to be near the stations due to data acquisition/equipment restraints and the position of nearby accelerometers. These being the `corner station', where the Beamsplitter, initial test masses and readout optics are, the `endstations' which contain the end test masses at the far end of each arm, and the `midstation' which sits in the middle of each beam tube, covering the gate valve positioned at \SI{2017}{\meter} from the Beamsplitter . There are $6$ accelerometers along each arm: one longitudinal and one transverse inside the midstation (\SI{2015}{\meter}), then exclusively longitudinal at \SI{150}{\meter}, \SI{1900}{\meter}, \SI{2100}{\meter}, and \SI{3950}{\meter} (see~Fig.~\ref{fig:AccDiagram}) \cite{nguyen2021, effler2015}. Where individual baffles could be identified inside and near the midstations, shakers were attached directly to these beam tube supports. The injections were focused on the central baffles inside and around the midstations only, and not at the ends of the beam tube. Preliminary injections made inside the corner station and endstations revealed that there is stronger coupling into the strain noise from the excitation of other structures inside the stations than of the beam tube baffles, and so the ends of each arm were not included in the testing campaign. The baffles chosen to inject at were the four around the middle of the arm, being baffles X105 to X108 and Y105 to Y108. We know their positions relative to the Beamsplitter reliably to within a meter \cite{BT_design}. See table \ref{tab:loc} for details.

\begin{table}[h]
\begin{tabular}{c|c}
    Baffle ID & Position, ${m}$ \\
    \hline
    X(Y)105   & 1982 \\
    X(Y)106   & 2015 \\
    X(Y)107   & 2019 \\
    X(Y)108   & 2053 \\
    
\end{tabular}
\caption{Table of baffles under investigation, with their approximate positions measured in meters from the Beamsplitter \cite{BT_design} Note that baffle positions are the same in both X- and Y-arms.}
\label{tab:loc}
\end{table}

%The baffles chosen were 105 (\SI{1982.2}{\meter}), 106 (\SI{2015.3}{\meter}), 107 (\SI{2018.6}{\meter}) and 108 (\SI{2052.3}{\meter}). 106 and 107 are accessed from inside the midstation, and 105 and 108 are accessed from inside the beam tubetunnel on either side. Accurate distance measurements were possible due to a monument on the floor of the midstation beneath the Gate Valve assembly, marking exactly \SI{2017}{\meter} along the beam path.

\begin{figure}
    \centering
    \includegraphics[width=\linewidth]{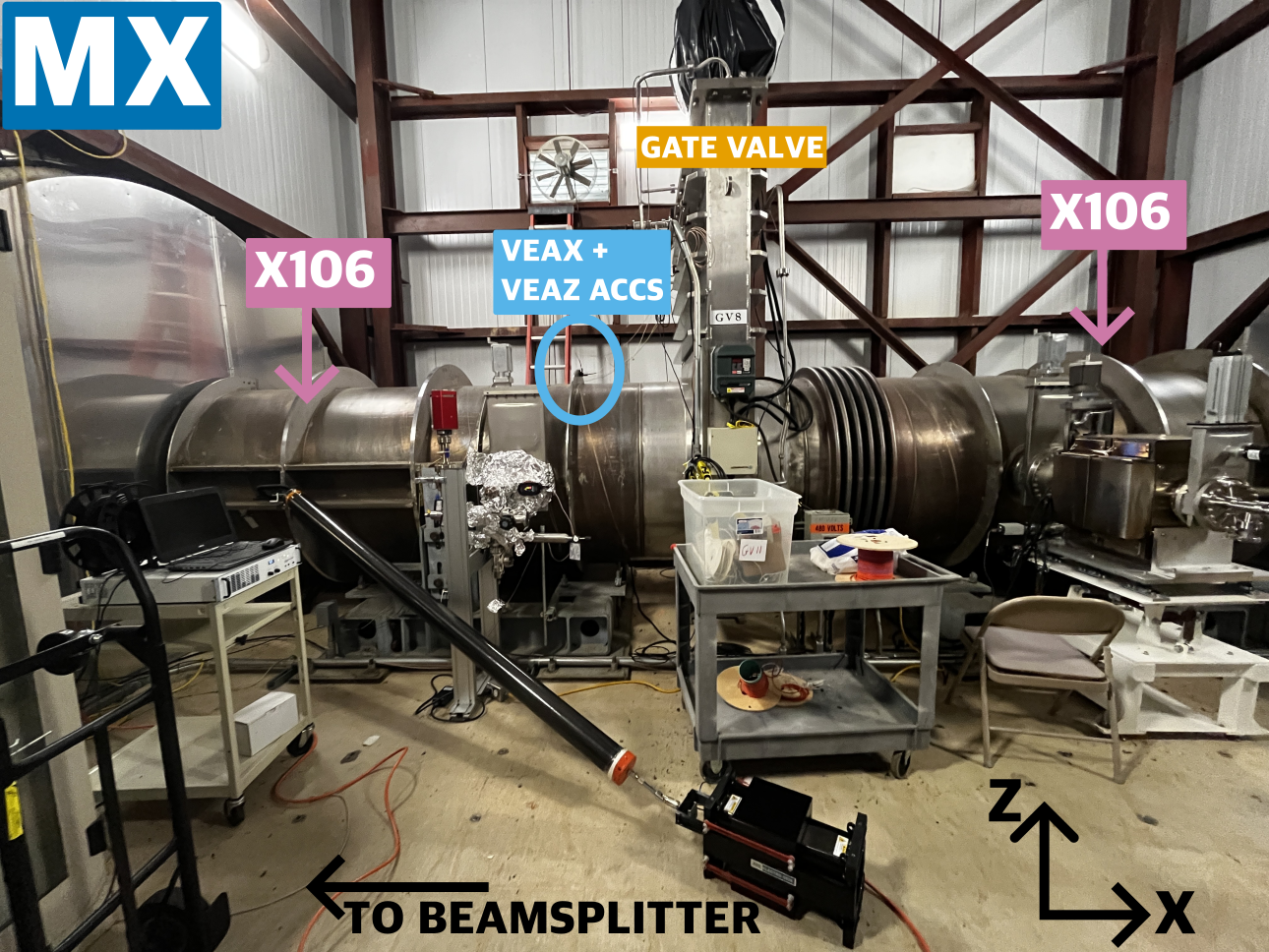}
    \caption{Image of a shaker injection at baffle X106 in the X-arm midstation. The locations of baffle X106 and X107 are noted, and the positions of accelerometers are also seen. The gate valve sits in the center above a `monument' which marks \SI{2017}{\meter} down the arm. An expansion bellow can also be seen to the right of the gate valve.}
    \label{fig:injpic}
\end{figure}

Injections focused on low frequencies, from \SI{5}{\hertz} - \SI{120}{\hertz} for the shaker and \SI{60}{\hertz} - \SI{200}{\hertz} for the acoustics. Each injection was broadband, with a typical width of \SI{20}{\hertz}, and lasted for \SI{128}{\second}. For the shaker tests specifically, the direction of motion was kept between \SI{30}{\degree} to \SI{40}{\degree} from the beam tube axis, in order to stimulate all three degrees of motion. The speaker used for the acoustic injections was positioned facing away from the midstation, so more of the beam tube would be directly exposed to the injection. The midstation tests had the speaker in the building facing the beam tube. To keep the injections as consistent as possible across all the injection sites, the nearest accelerometer trace was used and the injection power was tuned to be $100 \times$ above background motion before any measurements were taken. An example of a shaker injection inside a midstation is seen in figure \ref{fig:injpic}..

\subsection{Beam Tube Excitation} \label{sec:excitation}
\begin{figure*}[!t]
    \centering
    \begin{subfigure}{1\linewidth}
        \includegraphics[width=\linewidth]{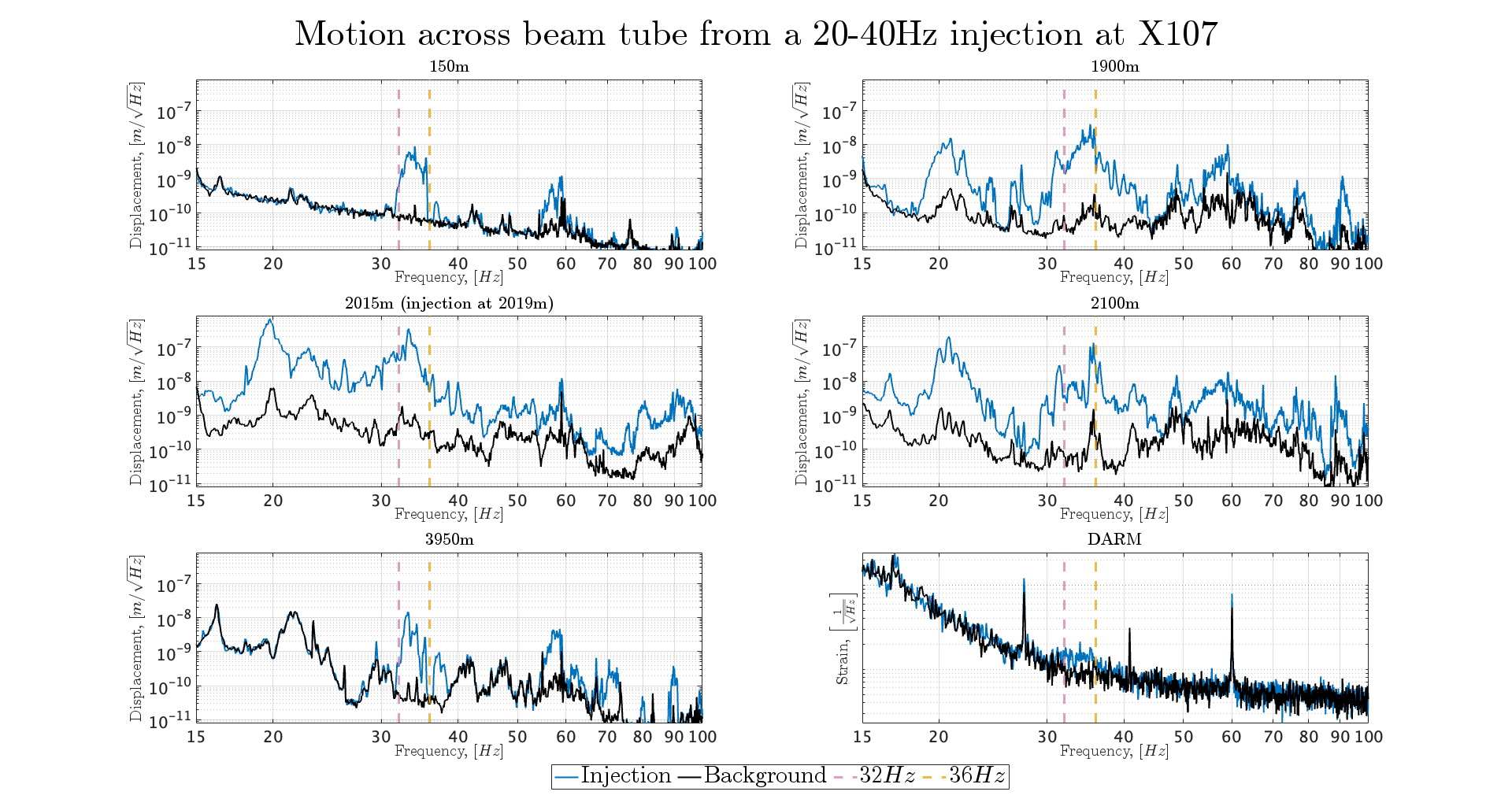}
        \caption{\SI{20}{\hertz}--\SI{40}{\hertz} injection at X107, showing \SI{32}{\hertz}--\SI{36}{\hertz} motion at both ends of the beamtube.}
        \label{fig:35allaccs}
    \end{subfigure}
    \begin{subfigure}{1\linewidth}
        \includegraphics[width=\linewidth]{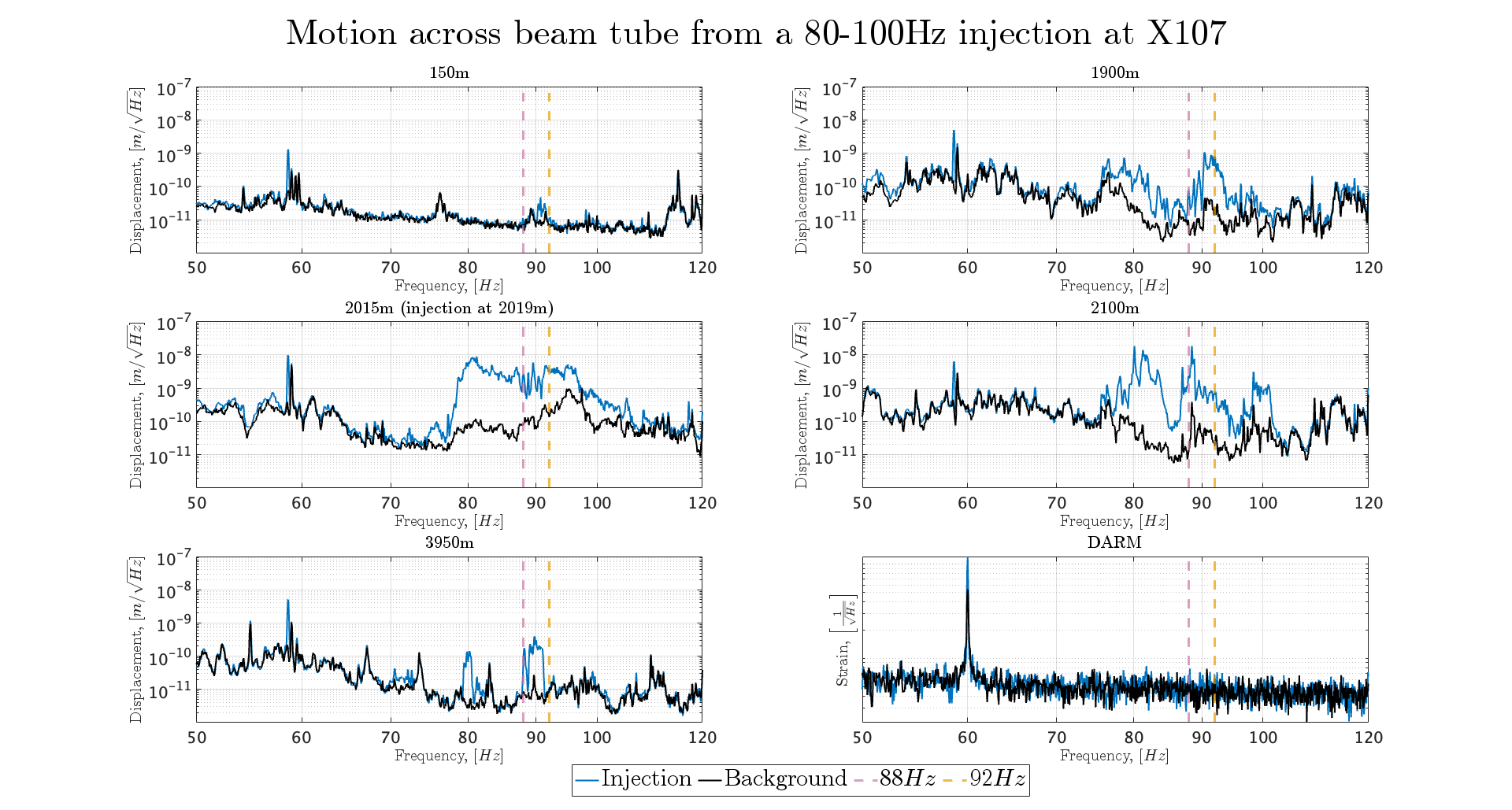}
        \caption{\SI{80}{\hertz}--\SI{100}{\hertz} injection at X107, showing \SI{88}{\hertz}--\SI{92}{\hertz} motion at both ends of the beamtube. Broadband seen around \SI{80}{\hertz} on far end of beam tube only, but no \SI{80}{\hertz} noise was seen in DARM for any tests.}
        \label{fig:90allaccs}
    \end{subfigure}
    \caption{Motion propagation along the beam tube for broadband injections at X107.}
    \label{fig:allaccs}
\end{figure*}

During injections we found that the X-arm was more responsive than the Y-arm. In total, we identified 13 tests where a broadband response was seen above the background in DARM, 8 of which were in the X-arm and 3 in the Y-arm. The tests revealed two broadband regions of interest, $\SI{32}{\hertz}-\SI{36}{\hertz}$ and $\SI{89}{\hertz}-\SI{91}{\hertz}$. Injected motion is not strongly attenuated in these two stated bands, with longitudinal beamtube motion at these frequencies decreasing by an order of magnitude  along \SI{2}{\kilo\meter} of the beam tube. The strength of propagation is evident in figure \ref{fig:allaccs}, in which \SI{20}{\hertz} bands were injected into the X-midstation, but only a few frequency bands propagated well to the ends of the beam tube. These regions also appeared in both arms at many injection sites.

% \begin{figure*}[tb]
%     \centering
    
%     \begin{subfigure}{.49\linewidth}
%         \centering
%         \includegraphics[width=\linewidth]{Images/Transfer2040.png}
%         \caption{Spectrogram of a \SI{20}{\hertz}-\SI{40}{\hertz} shaker injection inside midstation X at baffle X107}
%         \label{fig:34HzSpread}
%     \end{subfigure}
%     \hfill
%     \begin{subfigure}{.49\linewidth}
%         \centering
%         \includegraphics[width=\linewidth]{Images/Transfer80100.png}
%         \caption{Spectrogram of \SI{80}{\hertz}-\SI{100}{\hertz} shaker injection inside midstation X at baffle X106}
%         \label{fig:90HzSpread}
%     \end{subfigure}
%     \vfill
%     \begin{subfigure}{\linewidth}
%         \centering
%         \includegraphics[width=0.7 \linewidth]{Images/motion_on_bt_acs.png}
%         \caption{Accelerometer traces from \SI{2015}{\meter}, \SI{2100}{\meter} and \SI{3950}{\meter} along the arm, for the same midstation injection as in figure \ref{fig:34HzSpread}.}
%         \label{fig:AccelTraces}
%     \end{subfigure}
    
%     \caption{Spectrograms from across the X-arm, showing the $\approx$\SI{34}{\hertz} and $\approx$\SI{90}{\hertz} bands traveling across the beam tube.}
%     \label{fig:Spectrograms}
% \end{figure*}

In addition to these broadband features, of order 20 individual high-Q resonances were identified for each arm at the injection locations. On the X-arm, 6 of the high-Q features appeared in two separate injection sites, whereas each of the high-Q features in the Y-arm were localized to a single injection site and type. These resonances were all between \SI{60}{\hertz} and \SI{200}{\hertz}. See later sections and figure \ref{fig:HighQ} for a discussion of these high-Q features.

Another conclusion drawn from the data is that the gate valve in the midstations attenuates beam tube motion, due to the increased mass and support anchors at this point. This was discovered when comparing two identical shaker injections at baffles on either side of the gate valve which are $\approx 4$ meters apart. This same level of attenuation occurs across \SI{2}{\kilo\meter} of beam tube as occurs across the gate valve, so it follows that the large mass of the valve is strongly damping the motion of the beam tube. The tests also revealed that the air handling units located in and around the midstations can increase the accelerometer noise by a factor of a few, including in our well-propagating bands. This increased noise is not strong enough to appear in DARM with the current sensitivity, but it may become apparent as we move towards more sensitive detectors \cite{o2o3detchar, Capote2025, Soni2025}. To this end, air handling units that sit close to the beam tube should be carefully considered when designing next generation detectors, to ensure the characteristic peaks produced by these units do not couple into the well-propagating bands of the beam tube. 

\subsection{Stray Light Noise Modeling}
Calculating the expected stray-light noise in DARM involves modeling of the coupling with SIS and  measuring the motion of the beam tube and baffles (see sections~\ref{sec:model} and \ref{sec:excitation}). For the realistic scattering and propagation of the cavity field, modeling used measured surface mirror maps of the O4 test masses \cite{mirror_maps}. The BRDF at large scattering angles is not well described due to a limit of resolution on the mirror phase maps. This large scattering angle field has instead been simulated via an FFT analysis, informed by previous tests and the total losses measured in the arm cavities \cite{SIS}. This has been included in our simulations. 

Baffle backscattering noise is calculated using Eq.~\ref{eq:backscatter}. To understand the individual contributions of the baffles to this noise coupling, we isolate the terms in the equation that depend only on the baffle, and define a K-value:
\begin{align}
\label{eq:kvalues}
K_i = \left(\frac{1}{r_i}\right)^2\int_{baffle_i}\left(\frac{dP}{d\Omega_{ms}}\right)^2 \,d\Omega_{ms}
\end{align}
the K-value relates to the stray light amplitude A by: 

$$
 A = \sqrt{\lambda^2 \frac{dP}{d\Omega_{bs}} K}
$$

\begin{figure*}[!t]
    \centering
    \includegraphics[width=0.7\linewidth]{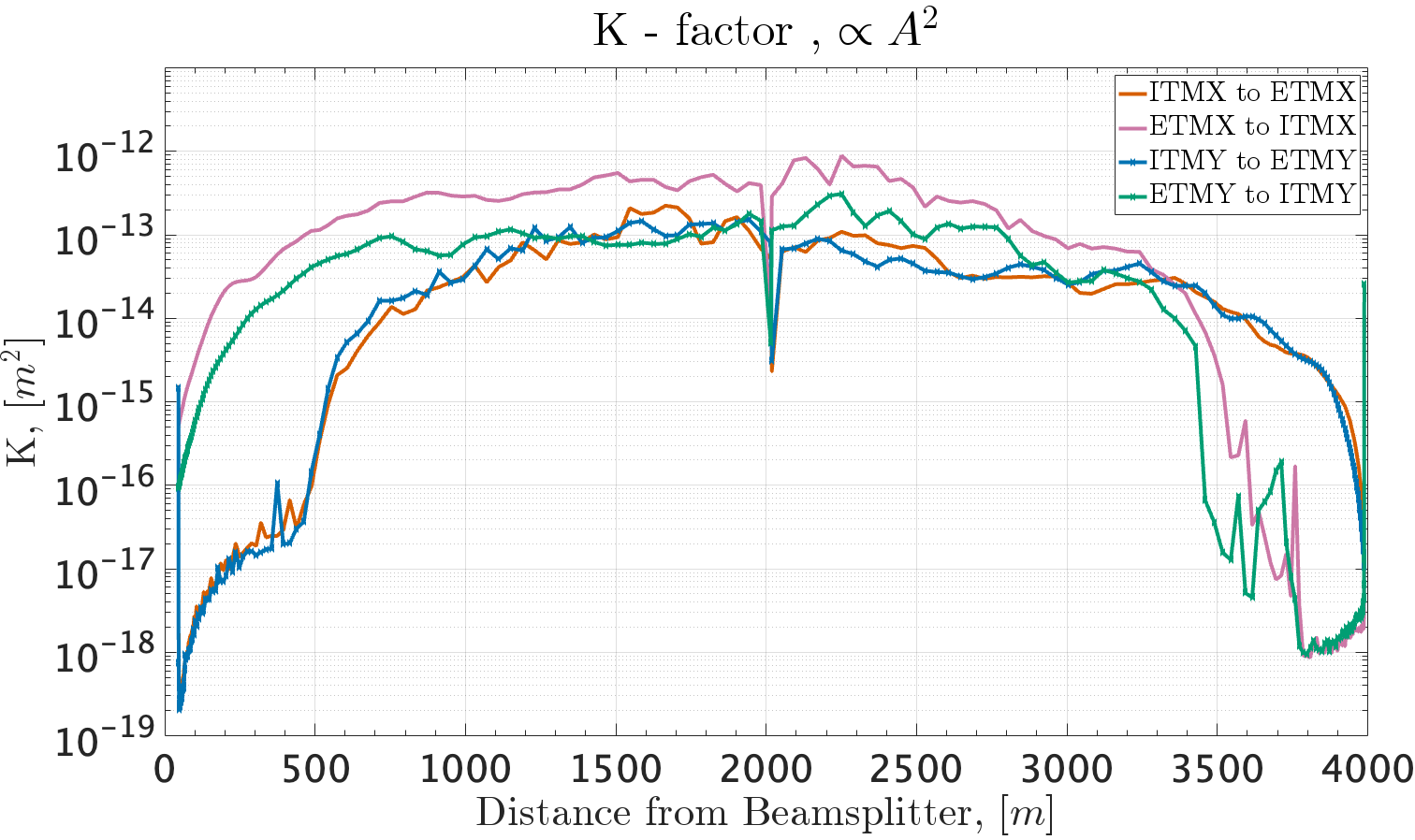}
    \caption{X-arm K-values (Eq.~\ref{eq:kvalues}) as calculated in SIS, for both ITM to ETM, and ETM to ITM. This data does not use baffle serrations.}
    \label{fig:Light_fraction}
\end{figure*}

As seen by this equation, the baffle K-values are dependent on both the separation of the baffle and test mass, and the power incident on it. The backscatter K-values for the X arm are shown in Fig.~\ref{fig:Light_fraction}. According to the model, the largest contribution to backscatter noise comes from baffles at the midpoint of the beam tube. The K-value is largest near the middle,  despite the large separation from the test masses, as the power scattered at lower angles is greater than at high angles. In addition, the baffle separation in the middle section is greater than at the ends of the cavity arms, and therefore there is less shielding from preceding baffles which contributes to the higher power incident on the baffle.
 
The crossover of the strain noise induced by the two backscatter components (phase noise and radiation pressure noise) is at approximately \SI{30}{\hertz}, which sits within the low frequency injection band.  This crossover point will shift slightly for each injection due to the motion spectrum used. Therefore it is important that both terms are considered in this band. 

Diffraction noise was calculated using Eq.~\ref{eq:diffraction}. While this noise coupling is estimated to be less important than backscatter noise, diffraction noise can be enhanced when high-Q baffle resonances are excited. Unlike backscatter noise where fringe wrapping broadens and flattens high-Q signals, diffraction preserves the shape of the resonance, which may in certain cases make the noise exceed that of backscatter. The diffraction noise arises from transverse motion, for which we do not have a sufficient number of accelerometers, so we will instead use the longitudinal motion as an approximation for the transverse motion. Upgrades planned for LLO may include adding transverse accelerometers along the beam tube, so this term may be better studied at a later date. 

% \subsubsection{Diffraction}
% We also consider the diffraction term in our summation. This is not the dominant noise contribution in the model for broadband injections, but is an important consideration in modelling the high-Q features. The diffraction noise arises from transverse motion, for which we do not have a sufficient number of accelerometers, so we will instead use the longitudinal motion as an approximation for the transverse motion. Upgrades planned for LLO may include adding transverse accelerometers along the beam tube, so this term may be better studied at a later date. We will use two separate SIS simulations, both with and without serrations for the diffraction and backscatter terms respectively. The pseudorandom serrations on the inner edges of baffles decrease the diffraction term, but do not affect the backscatter. The diffraction noise can be modelled as follows: 

% \begin{equation}
%     h(t) =  \frac{\lambda}{4 \pi L}\ |\Im(K)| \space  x(t)
% \label{eq:diff}
% \end{equation}
%  where $\Im(K)$ is the imaginary component of the SIS output. We then sum this term incoherently with the backscatter terms.
 
\subsection{Motion blending}
As SIS calculates K for each baffle individually, we need to estimate the motion for each individual baffle. We must also consider the effects of ground motion.  The accelerometers are less accurate at frequencies below \SI{5}{\hertz}, so we blend accelerometer and low-frequency seismic data to create a `supersensor' at each accelerometer position. One of the most prevalent noise sources at LLO is the microseism, where ocean waves impacting the continental shelf create low-frequency ground waves ($\approx \SI{0.03}{\hertz}$ to $\SI{0.1}{\hertz}$) \cite{Seismic_trains_scatter}. Microseism moves the ground surface in an elliptical motion both vertically and against the direction of travel. We use the vertical direction seismometer data as a proxy for longitudinal motion to ensure that the microseism noise is well-captured regardless as to the direction of travel, and very low-frequency ground tilt does not couple into this signal \cite{tilt2torsion}. The endstation STS-2 seismometers were chosen as there are no seismometers in the midstations and the corner station seismometer would have picked up more anthropic and equipment noise. These endstation seismometers sit on the ground a few meters from the BSC chambers that house the ETMs.

%The accelerometers are all mounted to the top of the beam tube, and so we create a supersensor blend with the seismic sensor inside the corner station, measured by a GS13. Though counter-intuitive, we use the Z-direction instead of X or Y. In the low frequency region, tilt motion couples into longitudinal signal but not into the Z signal. Therefore we get a cleaner approximation of low frequency longitudinal ground motion. The seismometer records in velocity, so we convert both the accelerometer and seismometer signals into displacement. The accelerometers are not as accurate at low frequencies, so we high-pass the accelerometers and low-pass the seismometer at \SI{7}{\hertz} and add them in the time domain. We then convert this supersensor signal to the Amplitude Spectral density (ASD) $\tilde{x}(f)$ for each accelerometer. 
 
As we do not have accelerometers at every baffle, we get the estimation for each individual baffle by performing a linear splice of known motion. We model the motion of each individual baffle as a linear combination of the two nearest supersensor traces. For example, a baffle $i$ at position $p_i$ from the Beamsplitter, which sits between the \SI{150}{\meter} and \SI{1900}{\meter} accelerometers would have a motion spectra of:

\begin{figure*}[!t]
    \centering
    \includegraphics[width=0.7\linewidth]{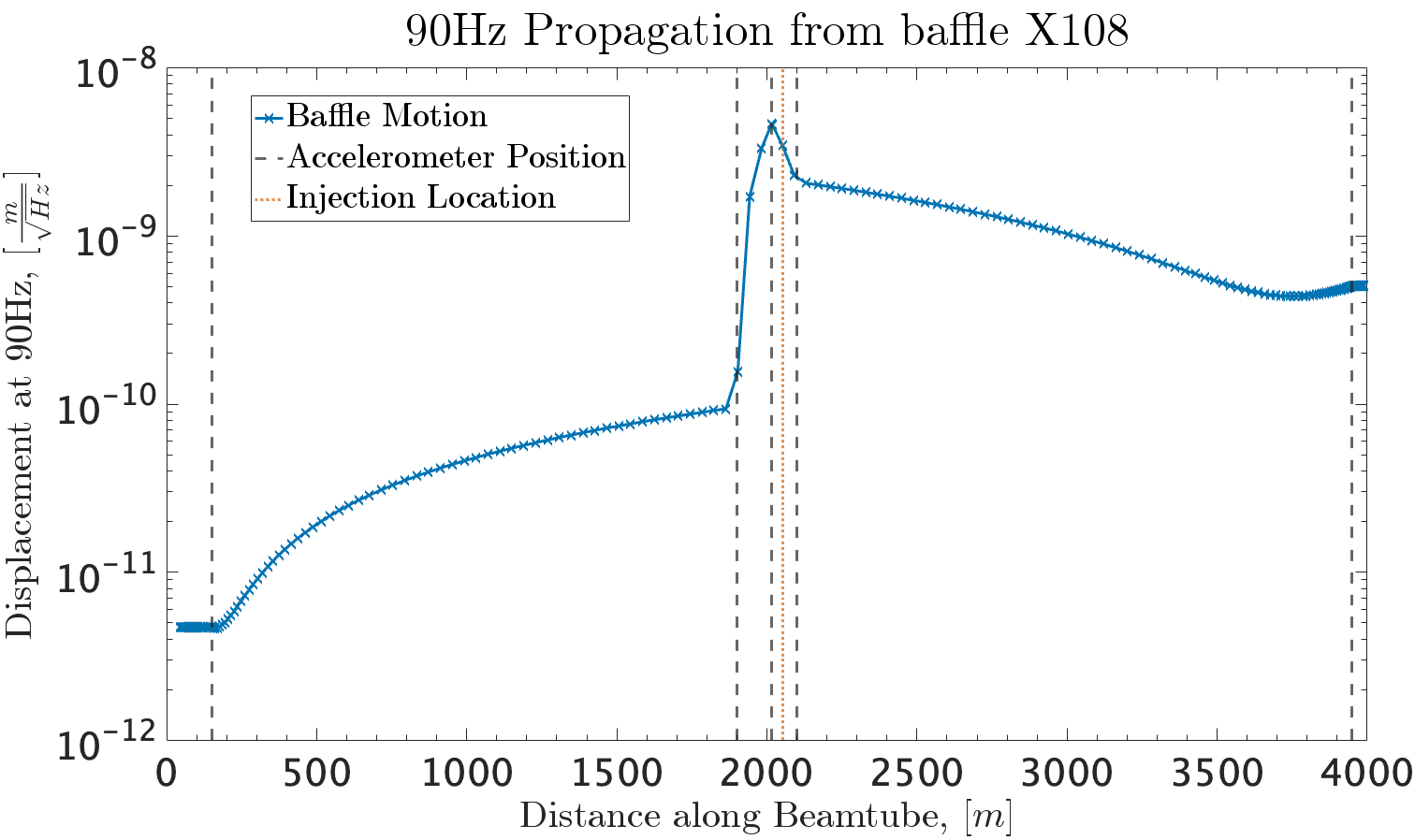}
    \caption{Spliced propagation of \SI{90}{\hertz} motion across the beam tube, following the method outlined in equation \ref{eq:interpolation}. The measured data here is from an \SI{80}{\hertz} - \SI{100}{\hertz} injection at X108. The positions of the accelerometers are also shown as dashed lines, and the location of baffle X108 is shown in the orange dotted line.}
    \label{fig:90TF}
\end{figure*}

\begin{equation}
\label{eq:interpolation}
\begin{split}
x_i(f) &=
\frac{p_i-150}{1900-150}\,x_{1900}(f) \\
&\quad +
\left(1-\frac{p_i-150}{1900-150}\right)x_{150}(f).
\end{split}
\end{equation}

 Baffles positioned under \SI{150}{\meter} and over \SI{3950}{\meter} we assume to have the same motion spectra as their respective end accelerometers. For example, figure \ref{fig:90TF} shows the estimated propagation of motion at \SI{90}{\hertz} from an injection at baffle X106 inside the midstation. The motion is calculated from the same interpolation detailed in equation \ref{eq:interpolation}. We again see the effect of the gate valve near the midpoint of the beam tube attenuating motion by approximately a factor of $100$.

Some preliminary work was undertaken to create a simple mass-spring model of the beamtube, in order to verify the spliced motion. This was unsuccessful, however. The `masses' were \SI{40}{\meter} tube sections with a mass of \SI{2200}{\kilo\gram}. The expansion bellows were modeled as the `springs' connecting the masses, and other `springs' modeled the the fixed supports connecting the tube to the ground. Recent measurement work performed by Coyne et.al. provided values of the axial spring constants of the fixed anchors and bellows \cite{Coyne}. The modeled propagation was attenuated much too strongly when compared to the measured propagation. Therefore this model was not used. Developing this model is beyond the scope of this project, so is left for future work. 

Now we have all of the components we can build the final model. We find the $\delta\tilde{h}$ by summing all three components (both backscatter components phase and radiation pressure), and diffraction incoherently in the frequency domain to find our total scatter, and then we can compare to the data for each successful test.
\FloatBarrier
\section{Results}\label{sec:results}
When comparing our simulations to the DARM spectra, we find that the predicted value of DARM is within an order of magnitude of observations across the two key frequency regions. This model is an incoherent summation of the background DARM trace, the diffraction and the backscatter. The full data can be seen in Table \ref{tab:data}. 

\begin{table*}
\begin{tabular}{|c|c|c|c|c|}
\hline
Location     & Injection [\si{\hertz}] & Type   & $\bar{\Delta}$ & $\Delta_{min}$ \\   \hline \hline
\multicolumn{1}{|c|}{X107} & \multicolumn{1}{c|}{20-40}                  & \multicolumn{1}{c|}{Shaker}  & \multicolumn{1}{c|}{7.9}  & \multicolumn{1}{c|}{4.2} \\ \hline
\multicolumn{1}{|c|}{X108} & \multicolumn{1}{c|}{32-36}                  & \multicolumn{1}{c|}{Shaker}  & \multicolumn{1}{c|}{9.0}  & \multicolumn{1}{c|}{2.4} \\ \hline
\multicolumn{1}{|c|}{X105} & \multicolumn{1}{c|}{32-36}                  & \multicolumn{1}{c|}{Shaker}  & \multicolumn{1}{c|}{4.1}  & \multicolumn{1}{c|}{2.0} \\ \hline
\multicolumn{1}{|c|}{X106} & \multicolumn{1}{c|}{80-100}                 & \multicolumn{1}{c|}{Shaker}  & \multicolumn{1}{c|}{10.4} & \multicolumn{1}{c|}{6.9} \\ \hline
\multicolumn{1}{|c|}{X105} & \multicolumn{1}{c|}{89-91}                  & \multicolumn{1}{c|}{Shaker}  & \multicolumn{1}{c|}{12.9} & \multicolumn{1}{c|}{4.8} \\ \hline
\multicolumn{1}{|c|}{X108} & \multicolumn{1}{c|}{80-100}                 & \multicolumn{1}{c|}{Shaker}  & \multicolumn{1}{c|}{5.6}  & \multicolumn{1}{c|}{2.8} \\ \hline
\multicolumn{1}{|c|}{X108} & \multicolumn{1}{c|}{89-91}                  & \multicolumn{1}{c|}{Shaker}  & \multicolumn{1}{c|}{8.2}  & \multicolumn{1}{c|}{7.1} \\ \hline
\multicolumn{1}{|c|}{Y107} & \multicolumn{1}{c|}{20-40}                  & \multicolumn{1}{c|}{Shaker}  & \multicolumn{1}{c|}{11.6} & \multicolumn{1}{c|}{7.3} \\ \hline
\multicolumn{1}{|c|}{Y108} & \multicolumn{1}{c|}{80-100}                 & \multicolumn{1}{c|}{Shaker}  & \multicolumn{1}{c|}{8.0}  & \multicolumn{1}{c|}{4.3} \\ \hline
\multicolumn{1}{|c|}{Y108} & \multicolumn{1}{c|}{80-100}                 & \multicolumn{1}{c|}{Speaker} & \multicolumn{1}{c|}{6.2}  & \multicolumn{1}{c|}{3.3} \\ \hline
MY               & 80-100                                      & Speaker                      & 8.6                       & 5.5                      \\ \hline
\end{tabular}

\caption{Final results from all key tests, showing the mean discrepancy between DARM measurement and model over a \SI{4}{\hertz} band (Either \SI{32}{\hertz} to \SI{36}{\hertz}, or \SI{88}{\hertz} to \SI{92}{\hertz}). The minimum discrepancy shows how close the model got to the measurements. The discrepancy is the factor by which the model is \emph{lower} than the measured strain. Note MY is the Y- midstation} 
\label{tab:data}
\end{table*}

\begin{figure*}[t]
    \centering
    \includegraphics[width=0.9\linewidth]{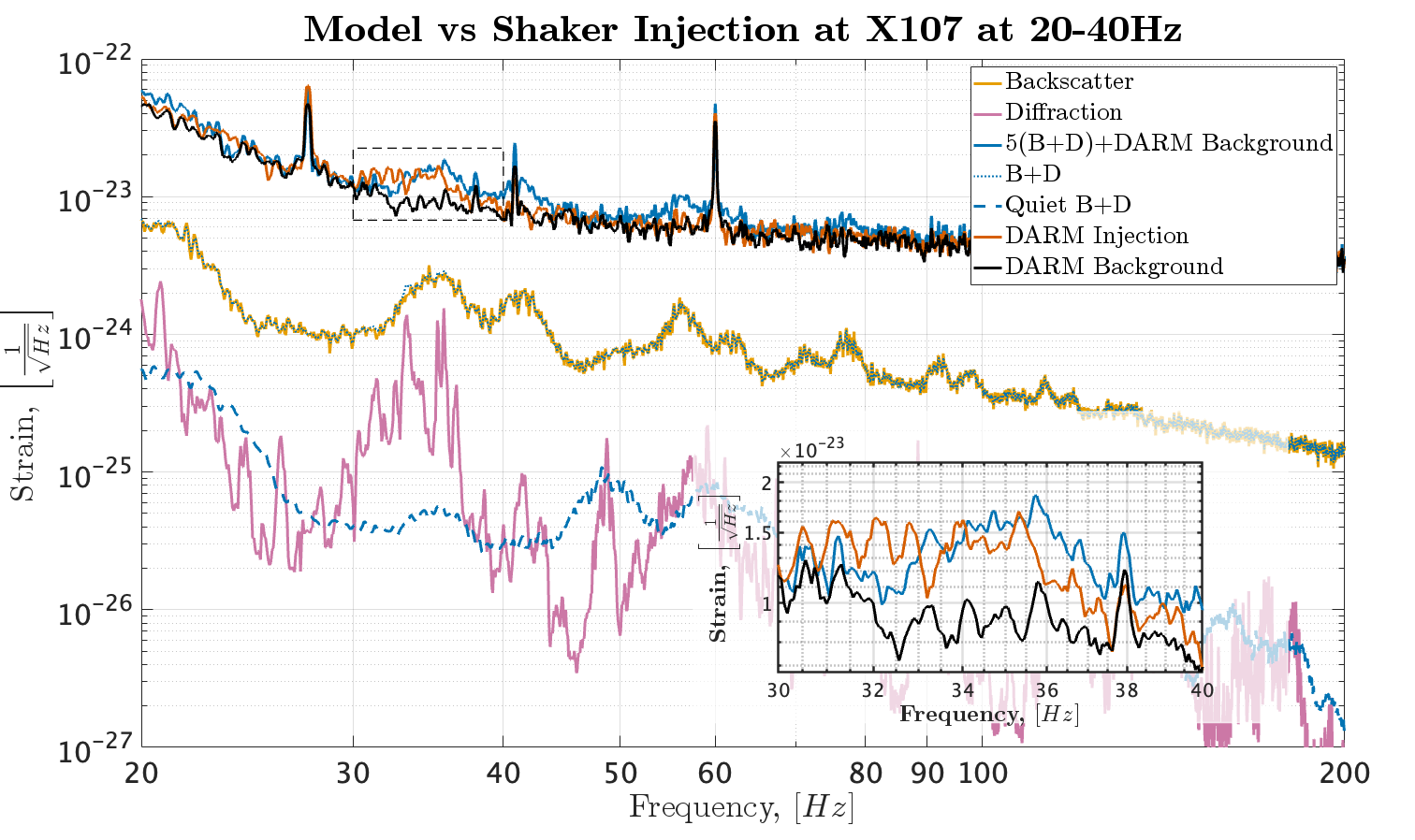}
    \caption{Model of a shaker injection at baffle X107 inside the midstation, with an injection of \SI{20}{\hertz} to \SI{40}{\hertz}.  Black: DARM background, red: DARM during the injection, yellow: backscatter noise, pink: diffraction noise. Blue is the combined model- blue solid is backscatter + diffraction + DARM background, multiplied by $\bar{\Delta}=5$, blue dotted is backscatter + diffraction with no multiplication factors applied, and blue dashed is the unmultiplied model with the background motion, not injected motion, to show the typical contribution of baffle stray light. The inset plot shows the region over which we see the broadband injection in DARM, here being \SI{30}{\hertz} to \SI{40}{\hertz}.}
    \label{fig:X-arm_midstation1072040}
\end{figure*}

In particular, the analysis highlights a measure of discrepancy between the DARM trace during injection and the model, termed $\bar{\Delta}$. This discrepancy is given by a factor that we multiply the model by to match the measured injection and minimize the discrepancy. This discrepancy is then averaged over a \SI{4}{\hertz} band (\SI{32}{\hertz} - \SI{36}{\hertz} or \SI{88}{\hertz} - \SI{92}{\hertz}). The minimum of this discrepancy factor within this band, $\Delta_{min}$,  is also found. Speaker injections were performed on the X-arm, but none of those tests showed a broadband response in DARM, so none are included. The location of the final speaker test is 'Y-midstation', as due to the size of the midstation and the proximity of the baffles inside the midstation, any sound will have affected both Y106 and Y107. For the test at Y108, the speaker was able to be positioned next to the baffle location.

The \SI{35}{\hertz} simulations had $\bar{\Delta} \approx 7$  below the DARM injection, while the \SI{90}{\hertz} region had $\bar{\Delta} \approx 10$.

Figure \ref{fig:X-arm_midstation1072040} shows the model vs injection for an X-arm test. This test in the \SI{20} - \SI{40}{\hertz} band, taken at X107 inside the midstation, matched the shape of the injection the most, showing that the broadening caused by fringe wrapping is apparent in the injections.

\begin{figure*}[t]
    \centering
    \includegraphics[width=0.9\linewidth]{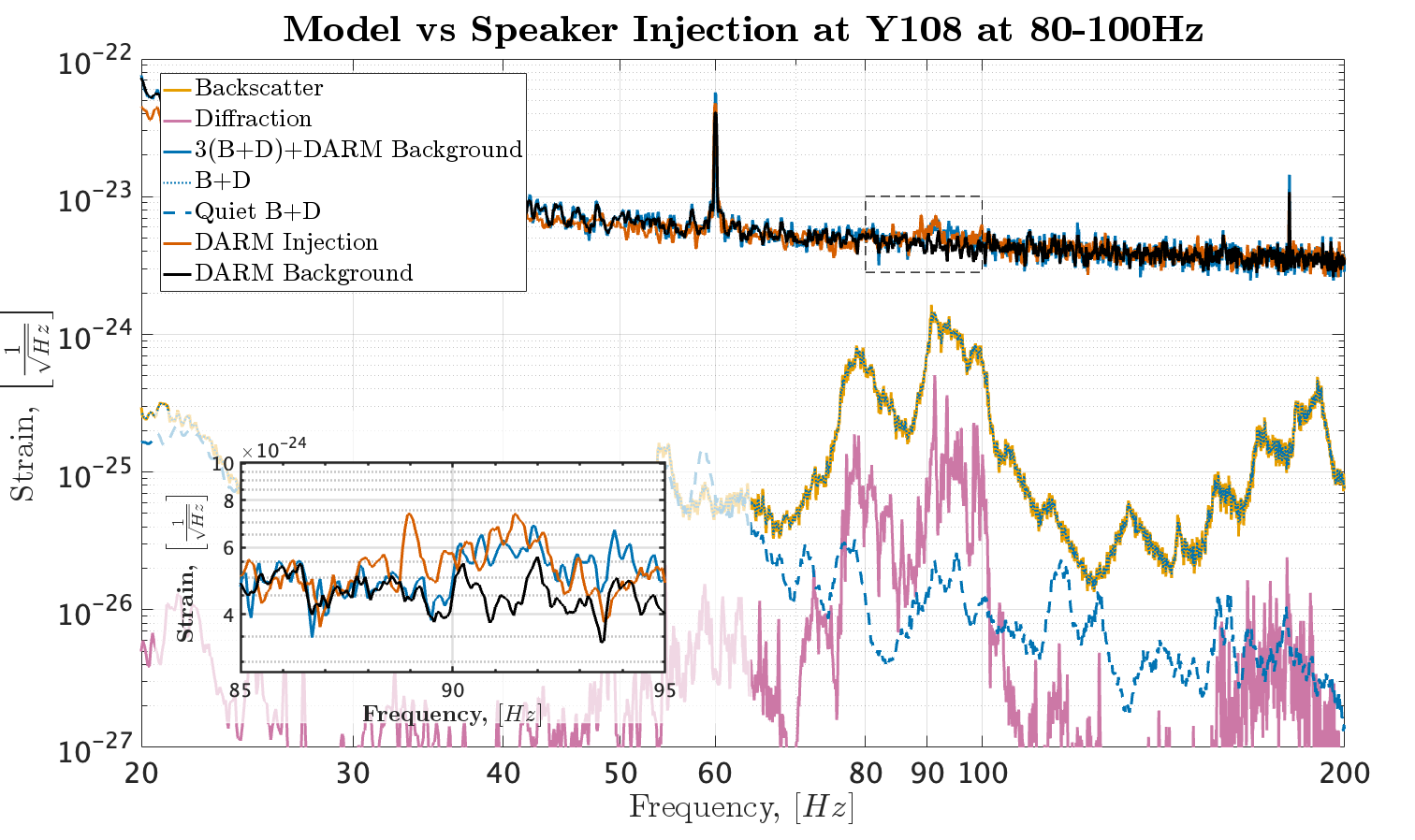}
    \caption{Model of a speaker injection at Y108, with an injected band of \SI{80}{\hertz} to \SI{100}{\hertz}.  See caption of figure \ref{fig:X-arm_midstation1072040} for details on traces. The inset plot shows the region over which we see the broadband injection in DARM, here being \SI{85}{\hertz} to \SI{95}{\hertz}.}
    \label{fig:Y90}
\end{figure*}

The Y-arm was not as reactive compared to the X arm, and so one test was found for the \SI{35}{\hertz}, and two tests for the \SI{90}{\hertz}. Conversely to the X arm, the three \SI{90}{\hertz} models were closer to the injection, with $\bar\Delta \approx 3 \mbox{ - }7$. The sole \SI{35}{\hertz} test had a $\bar\Delta =11.5$. Figure \ref{fig:Y90} shows the injection and model for Y108, at \SI{80}{\hertz}-\SI{100}{\hertz} via the speaker.

%We have also modelled the high-Q features discovered in and around the midstations. These features do not appear in the accelerometer traces, which shows that they may be as a result of the baffle itself resonating rather than the whole tube section. Sine waves with matching Q's as measured from the tests were added to the accelerometer traces in order to replicate the DARM features. After running the complete model, it is apparent that the DARM features are only a result of the diffraction noise. As described earlier, the diffraction noise is much more linear in displacement than the backscatter noise, and a strong injection (on the order of $0.001-1 \si{\milli\meter}$) would broaden out the backscatter via fringewrapping until most of the features are indistinguishable from the background. Therefore in subsequent models the sine waves were only added to the diffraction noise model. This gave a strong peak with no broadening. An example of the high-Q features can be seen in figure \ref{fig:HighQ} below. This figure also shows a \SI{90}{\hertz} broadband injection, which matches to a factor of 3 to 3.5. The high-Q features have been tuned and added separately, so they are not subjected to the multiplication factor needed to show the broadband features matching injection. Their magnitude ranged from \SI{1e-5}{\meter} to \SI{1e-3}{\meter}, with the peaks with the higher Q not needing as strong an injection compared to those with lower Q values.

\begin{figure*}[t]
    \centering
    %\includegraphics[width=0.9\linewidth]{Images/X2A_80100_fig_NicePlot.png}
    %\caption{High Q resonances and broadband inejction seen at X-2-A with an injected band of \SI{80}{\hertz} to \SI{100}{\hertz}. The peaks have measured frequencies and Q-factors as follows. \SI{81.6}{\hertz}: Q $\approx 400$,\SI{86.5}{\hertz}: Q $\approx 1500$, \SI{92.4}{\hertz}: Q $\approx 500$}.
    \includegraphics[width=0.9\linewidth]{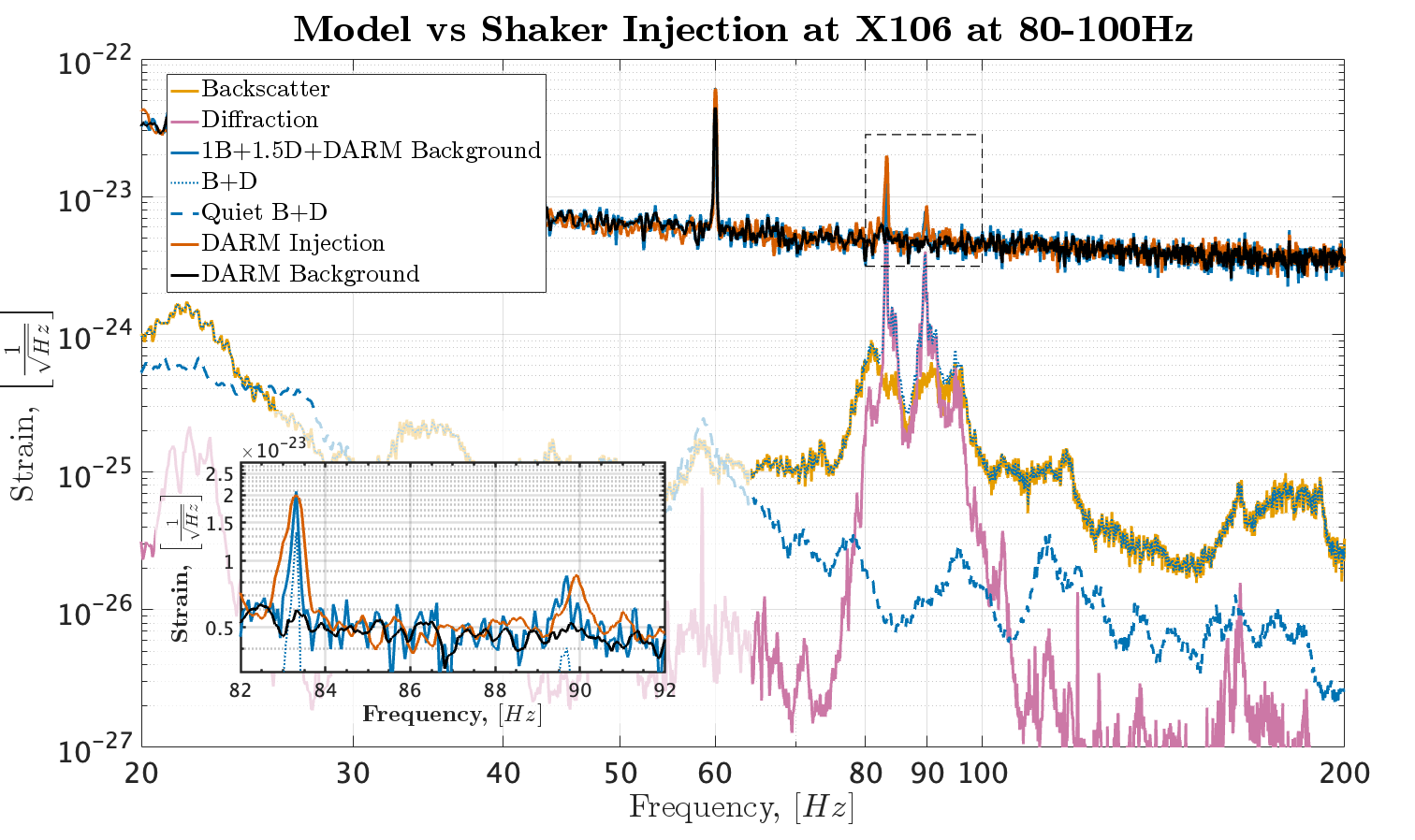}
    \caption{High-Q resonances and broadband injection seen at baffle 106 in X-arm midstation, with an injected band of \SI{80}{\hertz} to \SI{100}{\hertz}. The peaks, applied only to baffle X106, have measured frequencies and Q-factors as follows. $f = \SI{83.3}{\hertz}$: $Q=444$, $f = \SI{89.9}{\hertz}$: $Q=160$. See caption of figure \ref{fig:X-arm_midstation1072040} for details on traces.}
    \label{fig:HighQ}
\end{figure*}

We have also modeled the high-Q features discovered in and around the midstations.  These features are modeled to arise exclusively from diffraction noise pathways. As the rms motion of the baffles is much greater than the wavelength of the light, backscatter would cause harmonics in the strain signal which, when combined with upconverted low frequency motion like microseism, causes the peaks to broaden and flatten. This gives the characteristic appearance of fringewrapping. This does not, however,  occur in the much-more-linear diffraction noise, and so the fact that the peaks have remained sharp indicates that they are diffraction features \cite{T940063}. 

Additionally, these peaks did not appear in measured accelerometer traces, only in the strain signal. This could be that they were not strongly apparent in the few sections that include accelerometers, or they are stronger in the transverse direction which is largely unobserved due to only a single transverse accelerometer in the middle of the beamtube.

For the simplicity of modeling, we have \emph{assumed} that the high-Q resonances originate from from the resonant motion of individual baffles, and so we have applied the resonances to the fitted motion spectra of the baffle co-located with the injection. The remaining baffles use the approximated motion from the linear splice fit with no resonances applied. The high-q resonances were only applied to the motion when calculating the diffraction noise of the baffle, not the backscatter noise. The Q-factors of these resonances were found by measuring the full width at half maximum of the peaks in the DARM spectrum, and once applied to the baffle motion are in good agreement with the measured strain noise.

The origin of these resonances is currently unknown, and is left for future work. Correctly validating the origin of these high-Q features will require a wider variety and coverage of monitoring sensors are needed. For example, transverse accelerometers co-located with the longitudinal ones, and directional microphones spaced regularly and in both longitudinal and transverse orientations co-located with the accelerometers. One alternative to mechanical baffle resonances is that this noise may originate from transverse acoustic modes of the beam tube enclosure. As the inside of the enclosure is smooth concrete, low frequency sound from the shaker may propagate more than expected and create standing modes with similar Q-factors to those observed. Checking whether this noise is acoustic may be done by measuring the delay between two ends of the beamtube with an abrupt start of an injection. An example of one such model is shown in figure \ref{fig:HighQ}, which is a broadband injection at baffle 106 on the X arm. The high-Q resonance was only applied to the spliced motion of baffle X106 in this case.

%In order to test this, the calculated diffraction noise from the accelerometer trace is passed through a filter, an example of which is seen in figure \ref{fig:Filters}. This filter matches the test shown in figure \ref{fig:HighQ}, which is a \SI{80}{\hertz}-\SI{100}{\hertz} shaker injection at baffle 106 inside Mid-X. Once this filter is applied, the summation of twice the backscatter plus diffraction, plus background DARM matches the order of magnitude of the measured peaks. 

%\begin{figure*}[tb]
   % \centering
    %\includegraphics[width=0.9\linewidth]{Images/Filter_mx106_80100.png}
    %\caption{Filter used to replicate high-Q features. This is a summation of two second-order low-pass filters with the measured Q-factors. This is multiplied by the diffraction term, which gives the green trace in figure \ref{fig:HighQ}.}
%    \label{fig:Filters}
%\end{figure*}

%\millie{Note: changed the plot to use X-arm midstation106-80100 instead- two peaks, and a better match with $2*B + 1*D$ }
\FloatBarrier
\section{Conclusion}

This testing and modeling campaign at LIGO Livingston has verified that baffle motion can introduce stray light noise into the gravitational wave strain signal, and current stray light noise models are accurate to within a factor of approximately $3-10$.  Two key regions of well propagating  motion were identified, being around \SI{35}{\hertz} and \SI{90}{\hertz}. These propagate along the whole arm of the detector, and lead to broadband noise being seen in the strain signal. These features are is visible when injected baffle motion is approximately $100 \times$ above LLO's background motion. High-Q features in and around the \SI{90}{\hertz} frequency band. were also induced in the strain curve by the injections. These are expected to come from diffraction noise induced by resonances of individual baffles. The modeled high-Q features match the injected motion to within a factor of a few using Q-factors measured directly from the strain signal. Other key findings show that the midstation gate valve may be acting as a central node for beam tube motion, due to an approximately tenfold decrease in measured motion from opposite sides of the gate valve, and air handling units along the arms do induce noise in the well propagating bands. 

This work may be used to shape the baffle strategy of future detectors. For example,  if the high-Q resonances are found to be caused by individual baffle resonances, these can be reduced by addressing their installation and design to include damping. This could lead to a baffle similar to the suspended baffle design that is being considered for Cosmic Explorer \cite{CE_sus_Baff}. Understanding these high-Q features may also influence the enclosures and structures surrounding the beamtube. The well-propagating broadband noise may be mitigated by slightly randomizing the positions of the beam tube supports to suppress eigenmodes of the beam tube. Future stray light validation work will require a focus on more extensive measurement and modeling of beam tube and baffle motion, as instrumentation coverage has been a limiting factor during the testing. 

\section{Acknowledgments}

The authors would like to thank the LLO site team and LIGO fellows for their help during the measurement campaign. We would also like to thank Dennis Coyne and his team for sharing his data on beam tube support stiffness. This material is based upon work supported by NSF’s LIGO Laboratory which is a major facility fully funded by the National Science Foundation. LIGO was constructed by the California Institute of Technology and Massachusetts Institute of Technology with funding from the National Science Foundation, and operates under Cooperative Agreement PHY-1764464. Advanced LIGO was built under grant No. PHY-0823459. The authors are grateful for computational resources provided by the LIGO Laboratory and supported by National Science Foundation grants PHY-0757058 and PHY-0823459. EC and DM acknowledge the support of the Institute for Gravitational Wave Astronomy at the University of Birmingham, STFC Consolidated Grant “Astrophysics at the University of Birmingham” (No. ST/S000305/1), UKRI Quantum Technology for Fundamental Physics scheme (Grant No. ST/T006609/1 and ST/W006375/1), and UKRI “The next-generation gravitational-wave observatory network” project (Grant No. ST/Y00423X/1). AK acknowledges the support of the NSF award No. PHY-2308793.

\bibliography{mybibliography}

\end{document}